\documentclass[aps,pra,twocolumn,superscriptaddress]{revtex4-2}
\usepackage[encapsulated]{CJK}
\usepackage{graphicx}
\usepackage{float}
\usepackage{siunitx}
\usepackage{fixmath}
\usepackage{titlesec}
\usepackage{amsmath,amsfonts,amssymb}
\usepackage[colorlinks=true,linkcolor=blue,citecolor=blue]{hyperref}

\usepackage{amsmath}  \usepackage{amssymb}  \usepackage{amsfonts}  \usepackage{bm}  \usepackage{bbm}   \usepackage{bbold}   \usepackage{braket}  \usepackage{color}  \usepackage{comment}  \usepackage{dcolumn}  \usepackage{enumerate}  \usepackage{epsfig}  \usepackage{gensymb}  \usepackage{graphicx}  \usepackage{indentfirst}  \usepackage{lmodern}  \usepackage{mathrsfs}  \usepackage{mathtools}  \usepackage{psfrag}  \usepackage{pst-all}  \usepackage{soul}  \usepackage{units}  \usepackage{xcolor}

\def\EF{{E_{\rm F}}}     
\def\lamp{\lambda_{\rm p}}         
        
\begin{document} 

\title{Ultraconfined Plasmons in Atomically Thin Crystalline Silver Nanostructures}

\author{Vahagn~Mkhitaryan} 
\affiliation{ICFO-Institut de Ciencies Fotoniques, The Barcelona Institute of Science and Technology, 08860 Castelldefels (Barcelona), Spain}
\author{Andrew~P.~Weber} 
\affiliation{ICFO-Institut de Ciencies Fotoniques, The Barcelona Institute of Science and Technology, 08860 Castelldefels (Barcelona), Spain}
\affiliation{Donostia International Physics Center, Paseo Manuel Lardizabal 4, 20018 Donostia-San Sebasti\'an, Spain}
\author{Saad~Abdullah} 
\affiliation{ICFO-Institut de Ciencies Fotoniques, The Barcelona Institute of Science and Technology, 08860 Castelldefels (Barcelona), Spain}
\author{Laura~Fern\'andez} 
\affiliation{Centro de F\'{\i}sica de Materiales CSIC-UPV/EHU and Materials Physics Center, 20018 San Sebasti\'an, Spain}
\author{Zakaria~M.~Abd~El-Fattah} 
\affiliation{Physics Department, Faculty of Science, Al-Azhar University, Nasr City, E-11884 Cairo, Egypt}
\author{Ignacio Piquero-Zulaica} 
\affiliation{Centro de F\'{\i}sica de Materiales CSIC-UPV/EHU and Materials Physics Center, 20018 San Sebasti\'an, Spain}
\author{Hitesh~Agarwal} 
\affiliation{ICFO-Institut de Ciencies Fotoniques, The Barcelona Institute of Science and Technology, 08860 Castelldefels (Barcelona), Spain}
\author{Kevin~Garc\'{\i}a~D\'{\i}ez} 
\affiliation{Centro de F\'{\i}sica de Materiales CSIC-UPV/EHU and Materials Physics Center, 20018 San Sebasti\'an, Spain}
\author{Frederik~Schiller} 
\affiliation{Centro de F\'{\i}sica de Materiales CSIC-UPV/EHU and Materials Physics Center, 20018 San Sebasti\'an, Spain}
\author{J.~Enrique~Ortega} 
\email{enrique.ortega@ehu.es}
\affiliation{Donostia International Physics Center, Paseo Manuel Lardizabal 4, 20018 Donostia-San Sebasti\'an, Spain}
\affiliation{Centro de F\'{\i}sica de Materiales CSIC-UPV/EHU and Materials Physics Center, 20018 San Sebasti\'an, Spain}
\affiliation{Departamento de F\'{\i}sica Aplicada I, Universidad del Pa\'{\i}s Vasco, 20018 San Sebasti\'an, Spain}
\author{F.~Javier~Garc\'{\i}a~de~Abajo} 
\email{javier.garciadeabajo@nanophotonics.es}
\affiliation{ICFO-Institut de Ciencies Fotoniques, The Barcelona Institute of Science and Technology, 08860 Castelldefels (Barcelona), Spain}
\affiliation{ICREA-Instituci\'o Catalana de Recerca i Estudis Avan\c{c}ats, Passeig Llu\'{\i}s Companys 23, 08010 Barcelona, Spain}

\begin{abstract}
The ability to confine light down to atomic scales is critical for the development of applications in optoelectronics and optical sensing as well as for the exploration of nanoscale quantum phenomena. Plasmons in metallic nanostructures can achieve this type of confinement, although fabrication imperfections down to the subnanometer scale hinder actual developments. Here, we demonstrate narrow plasmons in atomically thin crystalline silver nanostructures fabricated by prepatterning silicon substrates and epitaxially depositing silver films of just a few atomic layers in thickness. Combined with on-demand lateral shaping, this procedure allows for an unprecedented control over optical field confinement in the near-infrared spectral region. Specifically, we observe fundamental and higher-order plasmons featuring extreme spatial confinement and high-quality factors that reflect the crystallinity of the metal. Our approach holds potential for the design and exploitation of atomic-scale nanoplasmonic devices in optoelectronics, sensing, and quantum-physics applications.
\end{abstract}
\maketitle
\date{\today}

\section{INTRODUCTION} \label{section:Introduction}

\noindent A large deal of work has been devoted to the confinement of light down to deep-subwavelength spatial regions as a powerful strategy to enhance nonlinear interactions with matter \cite{BBH03}, enable efficient single-molecule sensing \cite{NE97}, and explore quantum optics phenomena at the nanoscale \cite{AMY07}. Efforts have thus focused on localized and surface-propagating optical modes that are robust, long-lived, spatially compressed relative to free-space light, and lying within a spectral range of technological interest such as the near-infrared (NIR). Unfortunately, finding a single platform that combines these appealing properties is a hard task. Partial success has been obtained by engineering lossless dielectric structures \cite{ALC22}, although the mode sizes are yet far from the few-nanometer realm, particularly in the NIR. For example, polaritons in two-dimensional (2D) materials \cite{paper283,LCC17} (e.g., plasmons in graphene \cite{WLG15,NMS18} and phononic modes in hexagonal BN \cite{GDV18} and MoO$_3$ \cite{paper396}) are long-lived, extremely confined, and amenable to active tunability \cite{SMR23,paper404}, but they only exist at mid-infrared frequencies.

Atomic-scale confinement of NIR light requires hybridization with electronic excitations, with the ultimate limit of spatial compression being represented by individual atoms and molecules, although their application is in practice hindered by weak coupling to light compared with coupling to atomic vibrations in any host material unless cryogenic temperatures are introduced \cite{RWL12}. Analogously, excitons in transition metal dichalcogenides operate in the visible domain and are active at room temperature, but only within narrow spectral bands \cite{LCZ14,GLS19}.

Visible and NIR plasmons in noble metals are collective broadband electron excitations of bosonic nature, capable of strongly localizing optical energy down to the few-nanometer scale in all three spatial directions \cite{paper156}, and commonly displaying large transition strengths that are contributed by a large number of electrons. Through a sustained series of advances in colloid chemistry \cite{CM14} and nanolithography \cite{NLO09}, the plasmon mode frequency and spatial distribution can be tuned at will by adjusting the size and morphology of the supporting structure. Although inelastic losses limit the lifetimes of noble-metal plasmons down to tens of femtoseconds, their strong spatial confinement renders them as excellent nonlinear elements for harmonic generation \cite{BBH03,B08_3} and wave mixing \cite{HVQ12,ZWZ13}, as well as an optimum choice for optical sensing based on surface-enhanced Raman spectroscopy \cite{proc054}. In addition, the robustness and strong coupling to light displayed by noble-metal plasmons have been instrumental in the success of plasmonics in areas such as optical signal communications \cite{LHM13}, light modulation \cite{HCH18,HJD19,TZW20}, biosensing \cite{AHL08,KEP09,CA12_2,ORY21}, structural coloring \cite{KDH12,KYB16}, and control of thermal emission \cite{CLC15,YDC16}, in spite of the fact that most studies have relied on amorphous or polycrystalline samples. In this respect, it is well-known that defects and grain boundaries increase inelastic losses, and this effect can be dramatically reduced in single-crystal structures \cite{DHW05}.

Visible and NIR plasmons in 2D metals combine robustness with an intrinsically high degree of spatial confinement \cite{SRK17,SCR18,paper335,paper326} that makes them attractive as a platform for nonlinear nanophotonics \cite{paper382} and optoelectronics \cite{paper236}. Previously, we demonstrated plasmons in silver structures as thin as eight (111) atomic monolayers (MLs) on Si(111) with a high crystalline quality revealed by the measurement of sharp electronic quantum-well states and electron diffraction patterns, as well as a record-high surface conductance \cite{paper335}. Plasmons in 2D metals are characterized by short wavelengths compared to free-space light, and therefore, a source of momentum is required to excite them, such as that provided by external tips \cite{NPH95,CHM19} or through the evanescent field supplied by electron beams \cite{paper149}. However, we are generally interested in shaping the metal structure to gain control over the near field associated with plasmon guiding, localization, coupling to external light, and interaction with other structures. Electron beam lithography and focused ion beam are commonly employed to this end, and although these techniques are perfectly suited for patterning thick metal films, they produce damage and contamination in 2D materials. For example, the measured quality factor of ultrathin plasmons has been reported to be  $\sim4$ in $<2$~nm silver \cite{paper335} and $\sim1$ in $\approx3$~nm gold \cite{paper326}. Ultimately, the quality factor of deep-subwavelength plasmons is limited by intrinsic material properties, quantified in a Drude damping $\hbar\gamma\sim20$~meV in silver and $\sim70$~meV in gold \cite{JC1972}, such that, at the telecom wavelength of $\sim1550~$nm ($\hbar\omega=0.8$~eV photon energy), we have an upper bound of the quality factor $Q=\omega/\gamma\approx$~40 and 10 in silver and gold, respectively. In pursuit of these limits, alternative patterning methods that can minimize metal damage are highly desirable. 

Here, we observe plasmons in few-atomic-layer-thick crystalline silver nanostructures with an unprecedented combination of spatial confinement and lifetime. Our samples are made possible through the development of a fabrication method consisting in lithographically prepatterning the desired morphologies on a silicon wafer and subsequently evaporating silver films of epitaxial quality. We observe silver plasmons with quality factors approaching 10 in $\sim3$~nm-thick silver bowties, with the modes being laterally confined down to a $\sim10$~nm gap region. The flexibility of our fabrication method is illustrated by a wide range of NIR spectral features exhibited by particles of various designated shapes and sizes. In addition, we compare particle arrays with different periodicities and reveal radiative losses driven by inter-particle interactions even for well-separated structures. These results open a practical approach to high-quality NIR plasmonics on the few-nanometer scale, enabling applications in optical sensing, light modulation, and nonlinear nanophotonics by bridging the gap between traditional noble-metal nanostructures and 2D materials.

\section{RESULTS} \label{section:Discussion}

\begin{figure*}
\centering \includegraphics[width=1.0\textwidth]{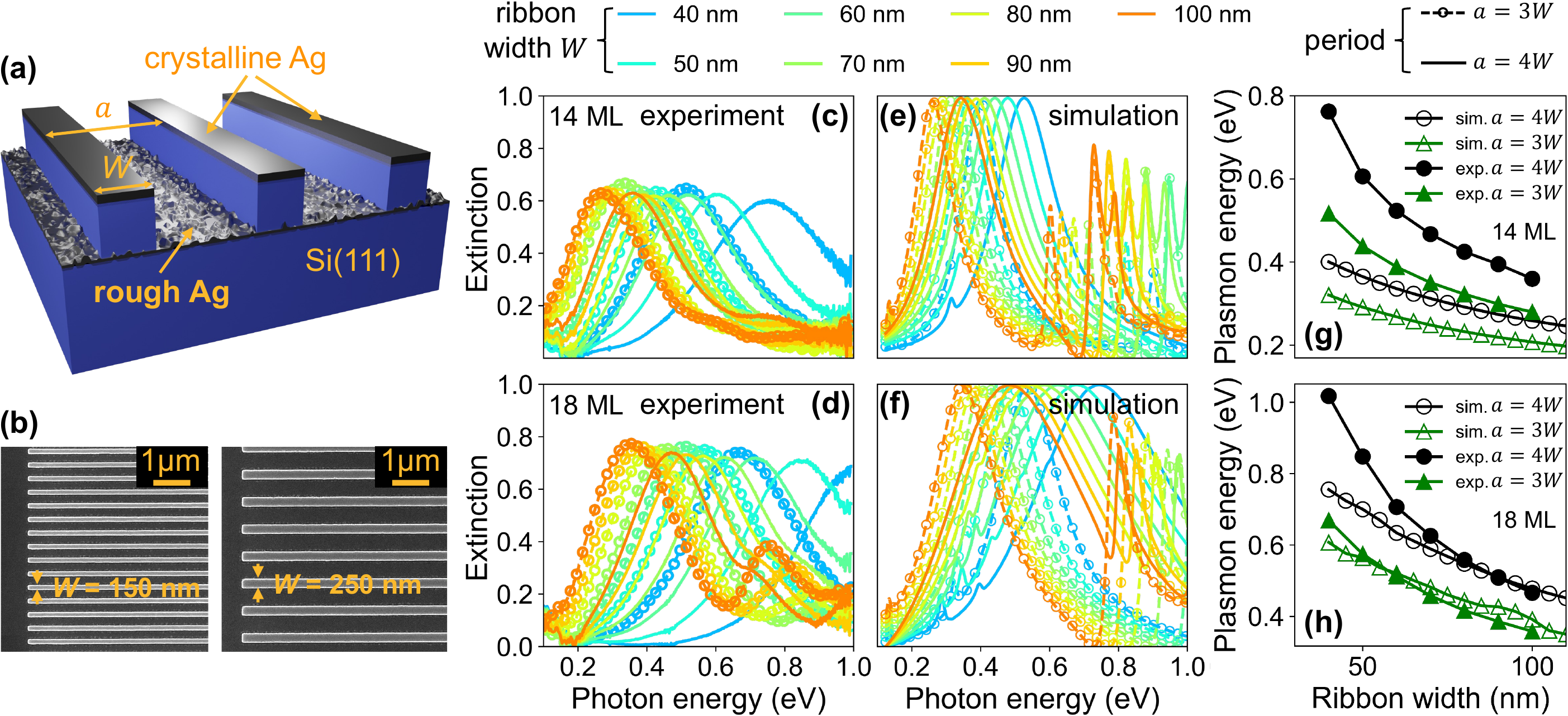}
\caption{{\bf Prepatterned silicon for high-quality ultrathin crystalline silver nanoplasmonics.} (a) Schematic representation of a ribbon array consisting of flat-top and rough-bottom Si(111) regions covered with an ultrathin Ag(111) film. (b) SEM images of representative samples with two different ribbon widths $W$ (see labels). (c,d) Measured optical extinction spectra for 14~ML~$\approx3.3$~nm (c) and 18~ML~$\approx4.3$~nm (d) silver films with different ribbon widths $W$ (see top legend) combined with an array period $a=3\,W$ (dashed curves with markers) or $a=4\,W$ (solid curves). (e,f) Simulations corresponding to panels (c,d), respectively. (g,h) Dipolar plasmon energies extracted from (c-f).}
\label{Fig1}
\end{figure*}

\noindent The sketch in Fig.~\ref{Fig1}a shows an example of our fabricated structures, consisting of silver ribbons deposited on a prepatterned silicon wafer and covered with a $\sim1$~nm passivating silica layer (not shown; see Methods). The lower regions in the structures are rough because we use a combination of chemical and physical dry etching, and accordingly, silver is also rough in those areas. In contrast, the upper silicon surfaces remain smooth because they are protected by a resist mask during etching (see also Supplementary Figs.~\ref{FigS1}-\ref{FigS4} for an extended surface-science analysis of our films). Examples of two fabricated ribbon arrays with widths $W=150$ and $250$~nm are shown in the scanning electron microscope (SEM) images of Fig.~\ref{Fig1}b. For ultrathin Ag(111) films consisting of 14 and 18~MLs ($\approx3.3$ and 4.3~nm, respectively), we collect extinction spectra that reveal well-defined plasmons over a wide range of ribbon widths $W=40-100$~nm (Fig.~\ref{Fig1}c,d). An expected redshift is observed as the ribbon aspect ratio is increased by either reducing the film thickness or making the width larger. In addition, the inter-ribbon separation plays an important role that we analyze by comparing structures with different periods: $a=3\,W$ (Fig.~\ref{Fig1}c,d dashed curves) and $a=4\,W$ (solid curves). In brief, shorter periods produce an attractive interaction between the modes of neighboring ribbons that translates into an observable redshift that becomes clearer when comparing the plasmon energies for the two period-to-width ratios under consideration (Fig.~\ref{Fig1}g,h). In addition, for a given plasmon energy, modes sustained by structures with a larger $a/W$ ratio are narrower, in agreement with the fact that the plasmon width receives a radiative contribution proportional to $\sim1/a$ \cite{paper335}.

We also show electromagnetic simulations of these structures based on a finite-difference method in the frequency domain (COMSOL), which are in excellent agreement with our measurements, as shown in Fig.~\ref{Fig1}e,f, particularly for the dipolar plasmon energies of the thicker sample (see Fig.~\ref{Fig1}h), although the experimental peaks are blue-shifted with respect to theory for the 14~ML films (Fig.~\ref{Fig1}g), presumably because of the effect of de-wetting in the rough metal deposited in the grooves at such small thicknesses, which changes its behavior from a continuous metallic layer to an effective dielectric composed of nonpercolated metal particles \cite{paper105}. The redshift produced by the attractive interaction of ribbon plasmons with the extended metal regions in the grooves is corroborated in our simulations (see Supplementary Fig.~\ref{FigS5}). We also observe an extra broadening in the experiment with respect to theory, which we attribute mainly to a finite distribution in the size of the fabricated ribbons.

Inelastic losses in the metal can also contribute to broadening, although the overall effect when contrasting experiments (Fig.~\ref{Fig1}c,d) and theory (Fig.~\ref{Fig1}e,f) is smaller for samples fabricated through the present prepatterning approach than what one finds when following a fabrication procedure based on lithography of previously deposited flat silver films \cite{paper335}. We thus understand that the high crystalline quality of the metal is better preserved by using the methods here reported. Supporting this conclusion, a decent preservation of silver quality is revealed by the observation of electronic quantum wells in angle-resolved photoemission spectroscopy (ARPES) measurements of patterned structures, comparable to those of extended unpatterned films (see Supplementary Figs.~\ref{FigS1} and \ref{FigS4}).

\begin{figure*}
\centering \includegraphics[width=1.0\textwidth]{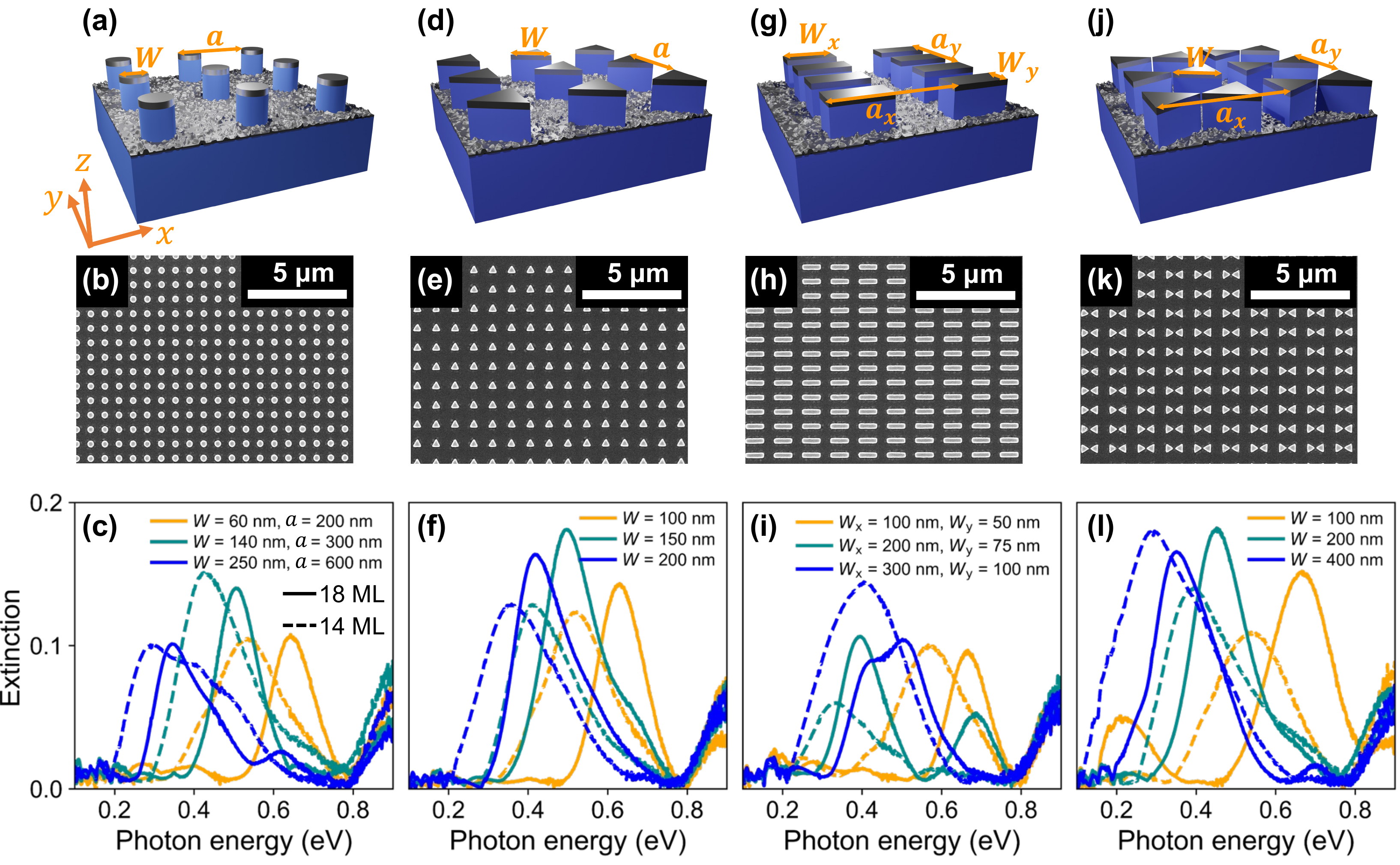}
\caption{{\bf Plasmon diversity in ultrathin crystalline silver nanostructures.} We study doubly periodic arrays of disks (a-c), triangles (d-f), bars (g-i), and bowtie antennas (j-l). The structures and geometrical parameters are illustrated in the top panels (a,d,g,j), accompanied by SEM images of representative fabricated samples (b,e,h,k). The measured extinction of these structures is shown in the bottom panels (c,f,i,l) for different lattice periods and structure sizes (see labels), combined with a silver film thickness of 14~ML (dashed curves) or 18~ML (solid curves). Disks and triangles are arranged in square lattices. Spacings for disks are indicated in the legend of (c). We set $a=2\,W$ for triangles (f), $a_x=2\,W_x$ and $a_y=3\,W_y$ for rods (i), and $a_{x}=3\,W$ and $a_{y}=2\,W$ for bowties (l). The gap in the latter is $\sim10$~nm.}
\label{Fig2}
\end{figure*}

The flexibility of the prepatterning approach allows us to fabricate a wide variety of ultrathin crystalline silver morphologies, whose plasmonic properties are analyzed in Fig.~\ref{Fig2}. We consider doubly periodic arrays of disks (Fig.~\ref{Fig2}a-c), triangles (d-f), bars (g-i), and bowtie antennas (j-l), whose geometrical parameters are defined in the upper sketches (Fig.~\ref{Fig2}a,d,g,j, respectively), along with SEM images of representative structures (Fig.~\ref{Fig2}b,e,h,k) that corroborate an excellent degree of quality. The corresponding extinction spectra (Fig.~\ref{Fig2}c,f,i,l) for silver thicknesses of 14~ML (dashed curves) and 18~ML (solid curves) display pronounced features associated with plasmon modes in the structures, revealing again a redshift in thinner structures relative to thicker ones. A change in morphology produces a radical redistribution in the position and arrangement of the observed spectral features. Interestingly, besides the main dipolar resonance peaks, additional modes are emerging in the upper energy range of the spectra under investigation, corresponding to higher-order plasmons that are more localized and, consequently, less radiative than dipolar plasmons. Such higher-order plasmons are thus spectrally narrower and exhibit larger quality factors, although, due to their weaker radiative coupling, their extinction cross sections are comparatively reduced. In addition, higher-order modes are associated with shorter plasmon wavelengths, which render them more sensitive to structural defects. The observation of high-quality, higher-order modes is thus clear evidence of the excellent quality of our fabricated silver structures.

\begin{figure*}
\centering \includegraphics[width=1.0\textwidth]{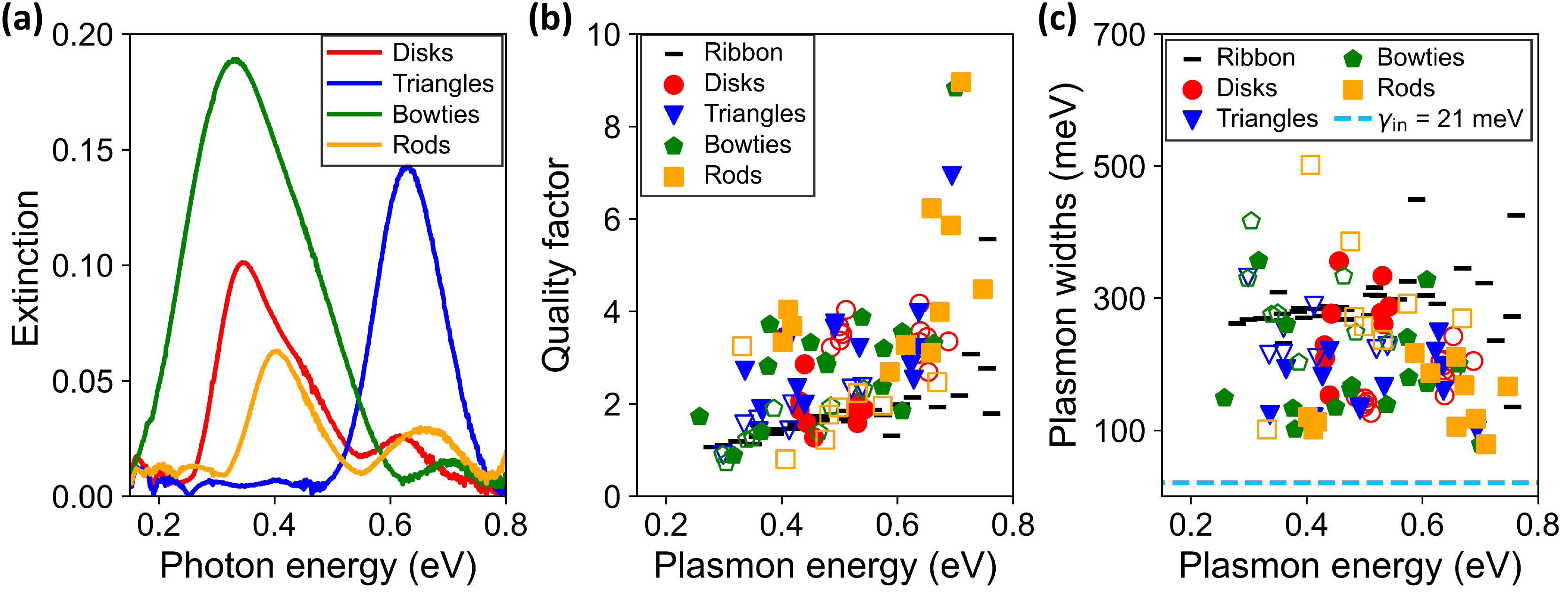}
\caption{{\bf Large quality factors in high-order plasmons.} (a) Selection of measured extinction spectra featuring high-order modes for different structure morphologies. (b,c) Quality factors (b) and plasmon widths (c) extracted from measured samples with different shapes (see legend) over the range of geometrical parameters considered in Fig.~\ref{Fig2} for a silver film thickness of 14~ML (open symbols) and 18~ML (solid symbols). Measuremed values for ribbon arrays are included for completeness. In (c), we show the intrinsic silver width of 21~meV (blue line).}
\label{Fig3}
\end{figure*}

Figure~\ref{Fig3}a shows a selection of spectra in different structures that sustain high-energy modes, from which we obtain the corresponding quality factor $Q_j=\omega_j/\gamma_j$ for each mode $j$ by fitting to a series of pseudo-Voigt line shapes and extracting the peak positions $\omega_j$ and resonance widths $\gamma_j$. Applying this procedure to a varied set of particle shapes and sizes, we obtain the quality factors and plasmon widths presented in Figs.~\ref{Fig3}b and \ref{Fig3}c, respectively. We find quality factors $Q>9$ in rods and bowties at relatively large energy ($\gtrsim0.7$~eV) in the NIR spectral range. A part of the observed plasmon broadening is directly inherited from the intrinsic optical Drude damping of the material, which is 21~meV \cite{JC1972} for bulk silver (blue line in Fig.s~\ref{Fig3}c). Additional damping is introduced by material defects, which act as an extra source of inelastic coupling from plasmons to electron-hole pairs. This effect is important when the structures are fabricated by patterning previously deposited metal films \cite{paper335}, but its magnitude is reduced for prepatterned structures in the present study, allowing us to reach substantially higher plasmon quality factors. Again, radiative damping contributes to the plasmon width when the structures are placed in close proximity, so that they are allowed to interact: for an individual structure (e.g., one of our ribbons or other shapes), coupling to radiation plays a negligible role because the lateral size is much smaller than the light wavelength; in contrast, when the distance between neighboring structures is reduced, the overall extinction takes substantial values (Fig.s~\ref{Fig3}a), and by reciprocity, coupling from plasmons to radiation is also becoming significant. For ribbon arrays, the contribution of radiative damping scales as the width-to-period ratio $\propto W/a$ \cite{paper335}, while for 2D arrays it goes as $\propto(W/a)^2$. The different scaling of ribbon and 2D particle arrays explains why, for the same metal thickness, the plasmon widths of ribbon arrays are generally larger than those of 2D arrays within the larger range of plasmon energies under consideration (Fig.~\ref{Fig3}c).

\begin{figure*}
\centering \includegraphics[width=1.0\textwidth]{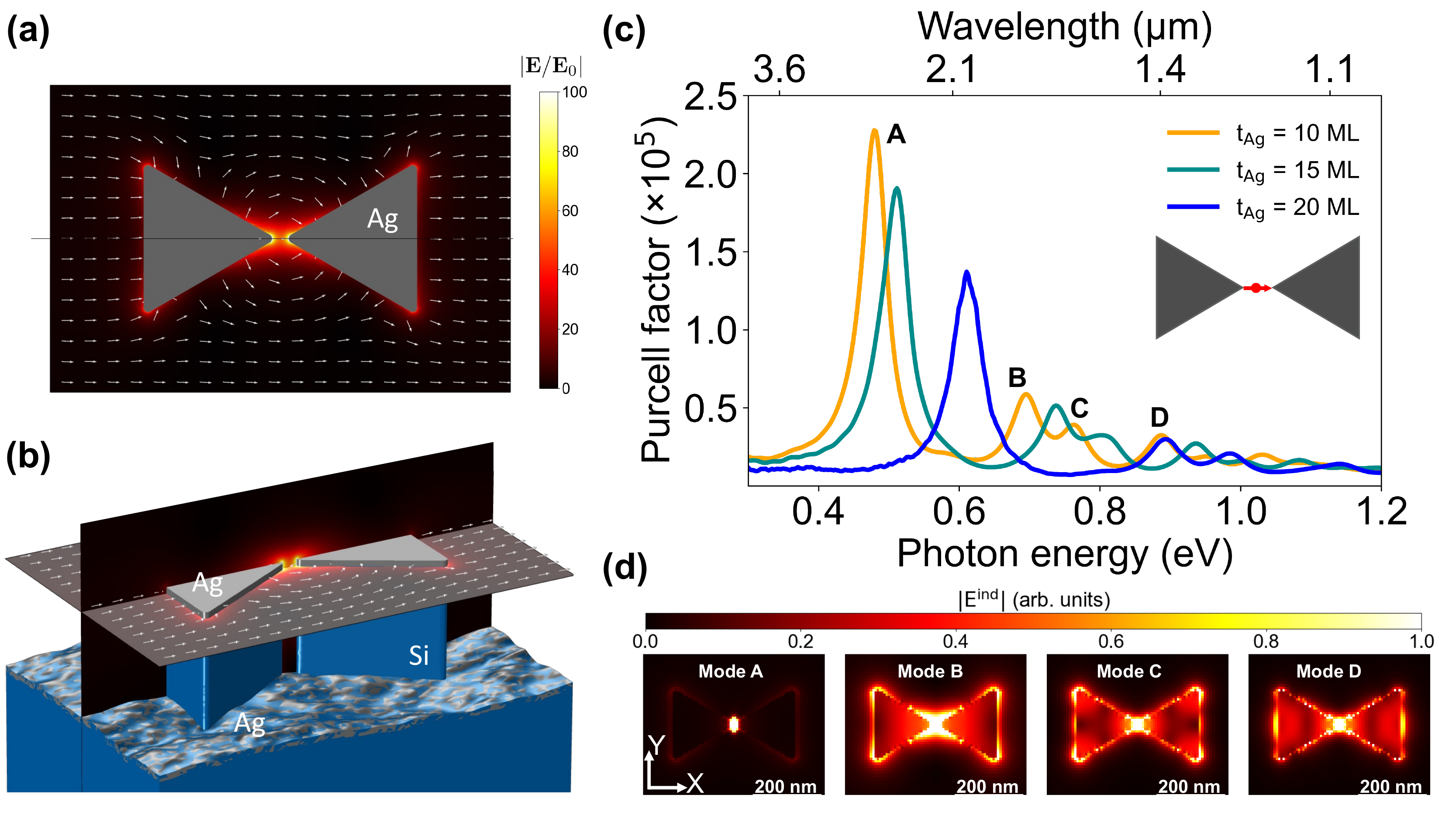}
\caption{{\bf Near-field distribution and Purcell factor simulated for bowtie structures.} (a,b) Electric near-field amplitude distribution for a silver bowtie of 18~ML thickness illuminated under normal incidence with light of 2.72~$\mu$m wavelength polarized along the gap direction. (c) Purcell factor for a dipolar emitter placed at the gap center of bowties with 10, 15, and 20~ML silver thickness. (d) Near-field distribution associated with features A-D in the 10~ML bowtie of (c). Triangles are equilateral with 200~nm side length and 10~nm gap separation in all cases.}
\label{Fig4}
\end{figure*}

Prepatterning of atomically thin films allows us to control not only the plasmon spectral properties but also the near-field distributions and the degree of localization of optical hotspots, such as in the bowtie structures presented in Fig.~\ref{Fig2}j-l. As an example, we theoretically explore in Fig.~\ref{Fig4} bowties made of equilateral triangles with $W=200$~nm side length. The near field associated with the lowest-order gap plasmon is illustrated in Fig.~\ref{Fig4}a,b for a 18~ML structure with a $\sim10$~nm gap at 2.72~$\mu$m light wavelength, showing that the field is primarily confined to the gap, with some intense features also emerging at distant corners of the structure, which can be attributed to the effect of charge redistribution guaranteeing neutrality in each of the two triangles. The gap-plasmon energy is very sensitive to geometrical details, such as the film thickness, with the associated Purcell factor (i.e., the decay rate of an emitter at the gap center relative to the rate in free space) taking high values of $\sim10^5$ within the fabricated range of thicknesses (Fig.~\ref{Fig4}c). Such high Purcell factors are the result of both the relatively narrow plasmon lines and the small volume contained in the gap regions. The calculated plasmon width is smaller than what we measure in the experiment (Fig.~\ref{Fig2}l), an observation that we attribute to the effect of averaging over a large number of structures in the spectrometer, which present unintended variations in geometrical parameters within the precision of the fabrication method. Incidentally, the bowtie structures also sustain higher-order modes with a small degree of field concentration in the gap compared to the fundamental gap mode (Fig.~\ref{Fig4}d).

\section{CONCLUSIONS}

\noindent The possibility of having damage-free, ultrathin, crystalline metal films marks a new upper bound in our ability to manipulate light at unprecedentedly small length scales. In this context, silver still holds a paramount position as a relatively low-loss plasmonic material, but its optical performance is limited by defects in noncrystalline samples, which introduce undesired inelastic losses. We demonstrate a low-damage fabrication method of ultrathin crystalline silver nanostructures with great flexibility to produce any desired lateral morphologies by epitaxially growing films with a controlled number of atomic layers on prepatterned silicon wafers. Using far-field near-infrared spectroscopy, we experimentally identify narrow plasmon modes supported in these structures with a quality factor approaching $Q\sim10$. We remark that high-order modes beyond dipolar features are also observed, thus increasing the level of spatial confinement down to a small fraction of the fabricated nanostructures. This is the case of bowties, which host gap modes with extreme spatial localization, while still retaining relatively high quality factors. Our results represent a substantial step towards the ultimate achievable level of quality factor for ultraconfined optical modes in the near-infrared spectral region, which is only limited by intrinsic metal losses in such deep-subwavelength plasmons (i.e., a Drude damping $\hbar\gamma=21$~meV in silver \cite{JC1972}, for which $Q=\omega/\gamma\sim40$ and 80 at 1550 and 800~nm light wavelength, respectively). At small particle sizes, plasmon confinement is further limited by nonlocal effects, which introduce additional damping mechanisms for sizes approaching $W\sim1$~nm \cite{paper119}, and thus, we envision the ultimate degree of combined confinement and lifetime in silver structures of $\sim10$~nm lateral size resonating at a light wavelength $\lambda_0\sim1$~$\mu$m with an in-plane plasmon wavelength $\lamp\sim2W$ and an out-of-plane extension $\sim\lamp/2\pi$ \cite{paper335}, resulting in a mode volume $\sim10^{-7}\,\lambda_0^3$. The present work represents a solid step in that direction, holding potential for the realization of sought-after nanoplasmonic applications in optoelectronics, optical sensing, and the exploration of quantum-optics phenomena at the few-nanometer length scale.

\section*{METHODS} \label{sec:Methods}

{\it Sample Preparation.} Silicon chips were prepatterned through a combination of electron-beam lithography (EBL) and dry reactive ion etching (RIE). Samples were first cleaned by ultrasound sonication in acetone and isopropanol for 5-10~min in each solvent, followed by blow-drying using nitrogen gas, and subsequent baking for 5~min at 150\textdegree{C} on a hot plate to remove any remaining water before a negative-tone AR-N-7520 EBL resist was spin-coated at 4000~rpm. The desired patterns were transferred to the resist using a Crestec lithography setup, followed by development in AR-300 for 1~min and rinsing in de-ionized water for 30~s. The patterned resist then served as a mask for etching the silicon in a RIE Oxford Plasmalab 80 Plus system employing a mixture of C$_4$F$_8$ and SF$_6$. With gas flows of 90~sccm for C$_4$F$_8$ and 30~sccm for SF$_6$, combined with a fixed radiofrequency power of 10~W and a coil power of 600~W, we obtained an etching rate of 80~nm/min as well as rather straight side wall profiles in the etched silicon structures. Etched samples were then washed in reagent-grade acetone and isopropanol for $>15$~min each. The native silica surface layer was removed in a commercial (iDonus) vapor-based etching system, using a solution of 40\% HF (by mass) in de-ionized water as the etchant, or by direct immersion in 2\% aqueous HF solution. Samples were etched for either 1~hr or $\approx90$~s when following vapor- or solution-based approaches, respectively, and then rinsed twice by immersion in de-ionized water for 5~min, dried under a stream of nitrogen, loaded into the ultrahigh vacuum (UHV) chamber within 12~min, degassed at a temperature in the $150-250$\textdegree{C} range for several hours, and heated to $>680$\textdegree{C} for over 2~min until a well-ordered Si(111)$-(7\times7)$ surface was identified by low-energy electron diffraction (LEED, see Supplementary Fig.~\ref{FigS2}). Silver was then sublimated from an electron-bombardment evaporator onto the Si(111)$-(7\times7)$ surface. The deposition, which we monitored with monolayer accuracy using a quartz microbalance whose response to a given silver flux was calibrated by probing the quantum-well state binding energies of silver films grown on Au(111) \cite{FGN11}, was carried out at a rate of $\approx0.4$~ML/min with the sample kept at a temperature of 110~K. Samples were then annealed to room temperature and the film thickness and quality were verified by photoemission measurements, followed by deposition of 1~nm of Si before removal from the UHV chamber, when oxidation transformed it into a passivating silica layer \cite{paper335}.

{\it Sample Characterization.} ARPES was performed using a SPECS Phoibos 150 electron analyzer equipped with a monochromatized He gas discharge lamp operating at the He I$\alpha$ excitation energy (21.22~eV), with an electron energy and angular resolution better than 40~meV and 0.4\textdegree, respectively. ARPES measurements (see Supplementary Fig.~\ref{FigS1}) were made with the sample at 110~K. X-ray photoelectron spectroscopy was also collected using a monochromatized Al K$\alpha$ source (1486.71~eV) to corroborate a low degree of carbon and oxygen surface contamination (see Supplementary Fig.~\ref{FigS3}).

{\it Optical Measurements and Analysis.} Plasmon spectra were acquired in the far field using a Bruker Hyperion Fourier-transform infrared (FTIR) spectrometer operating over a spectral range of $0.074-0.99$~eV. A multi-peak procedure was followed to assimilate the measured FTIR spectra to a combination of asymmetric pseudo-Voigt peak profiles. Plasmon resonance positions $\omega$ and FWHM $\Delta\omega$ were then extracted, from which the quality factors were calculated as $Q=\omega/\Delta\omega$.

\renewcommand{\thefigure}{S\arabic{figure}}

\begin{figure*}[h]
\begin{centering} \includegraphics[width=0.8\textwidth]{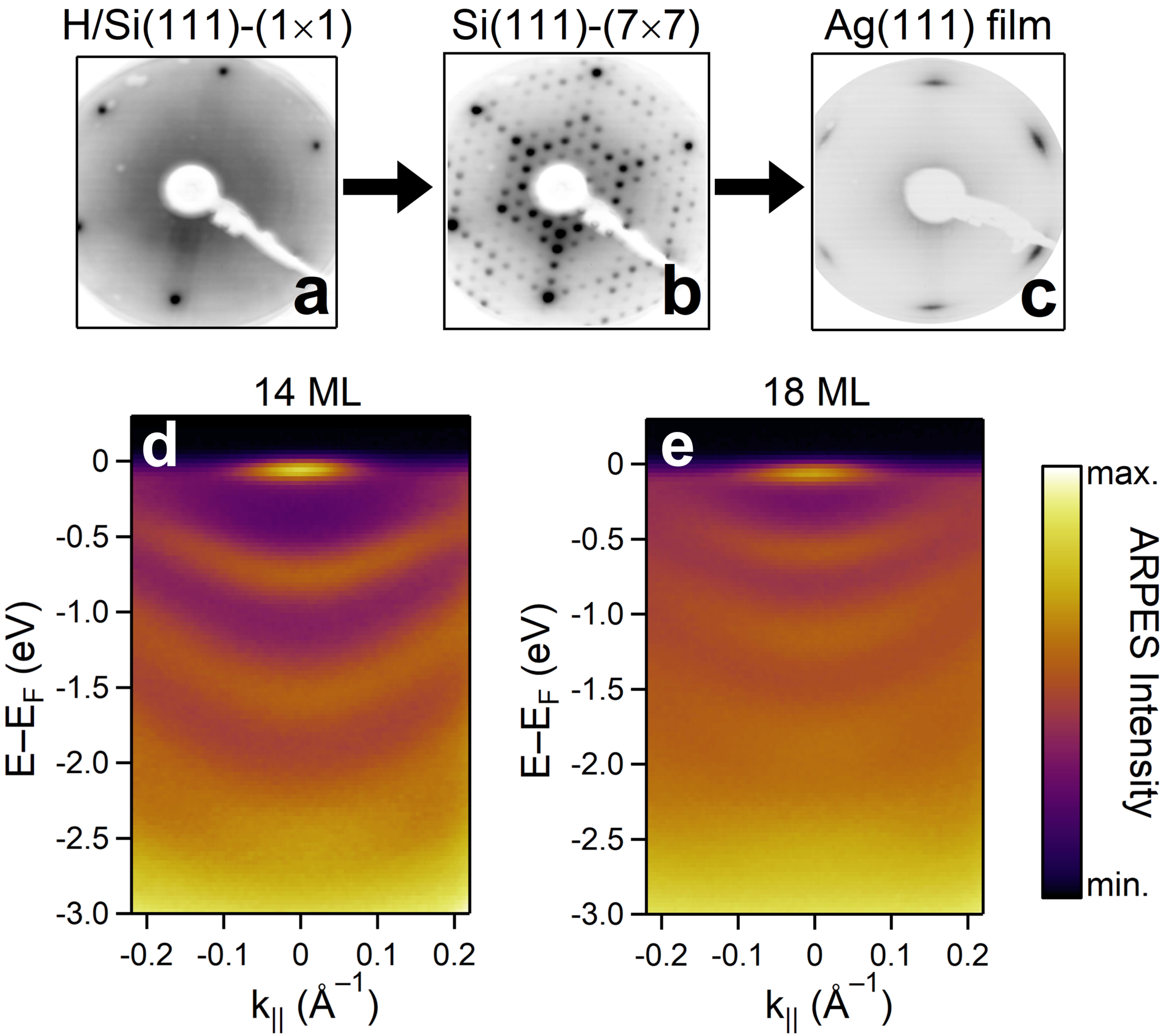} \par\end{centering}
\caption{Characterization of the silver film quality on H-passivated prepatterned Si(111). (a-c)~Low-energy electron diffraction (LEED) images providing a surface structural characterization of the procedure for preparing silver films on H-passivated prepatterned Si(111) in ultrahigh vacuum. The H/Si(111) surface presents $1\times1$ order until hydrogen desorption is induced by heating to a temperature above 550~$^{\circ}$C, leaving a bare Si(111)$-(7\times7)$ surface \cite{VES94}. Although our H/Si(111) surface shows sharp LEED spots in (a), there is a moderate background due to diffuse (and possibly inelastic) scattering by surface contaminants. To produce a clean Si(111)$-(7\times7)$ surface for high-quality silver epitaxy, the samples were gradually heated and kept at the maximum temperature of $650$~$^{\circ}$C for $12-15$~min. Following the procedure in Ref.~\cite{paper335}, the samples were then cooled below 120~K for silver deposition.~(d-e)~Angle-resolved photoelectron spectroscopy (ARPES) measurements showing the photoemission intensity as a function of binding energy relative to the Fermi level $\EF$ and in-plane wave vector for forms consisting of 14~ML (d) and 18~ML (e) of Ag(111) deposited on prepatterned silicon. The incident electron beam energy is 32~eV in (a,b) and 51~eV in (c). The ARPES photon energy is 21.22~eV.}
\label{FigS1}
\end{figure*}

\begin{figure*}
\begin{centering} \includegraphics[width=0.85\textwidth]{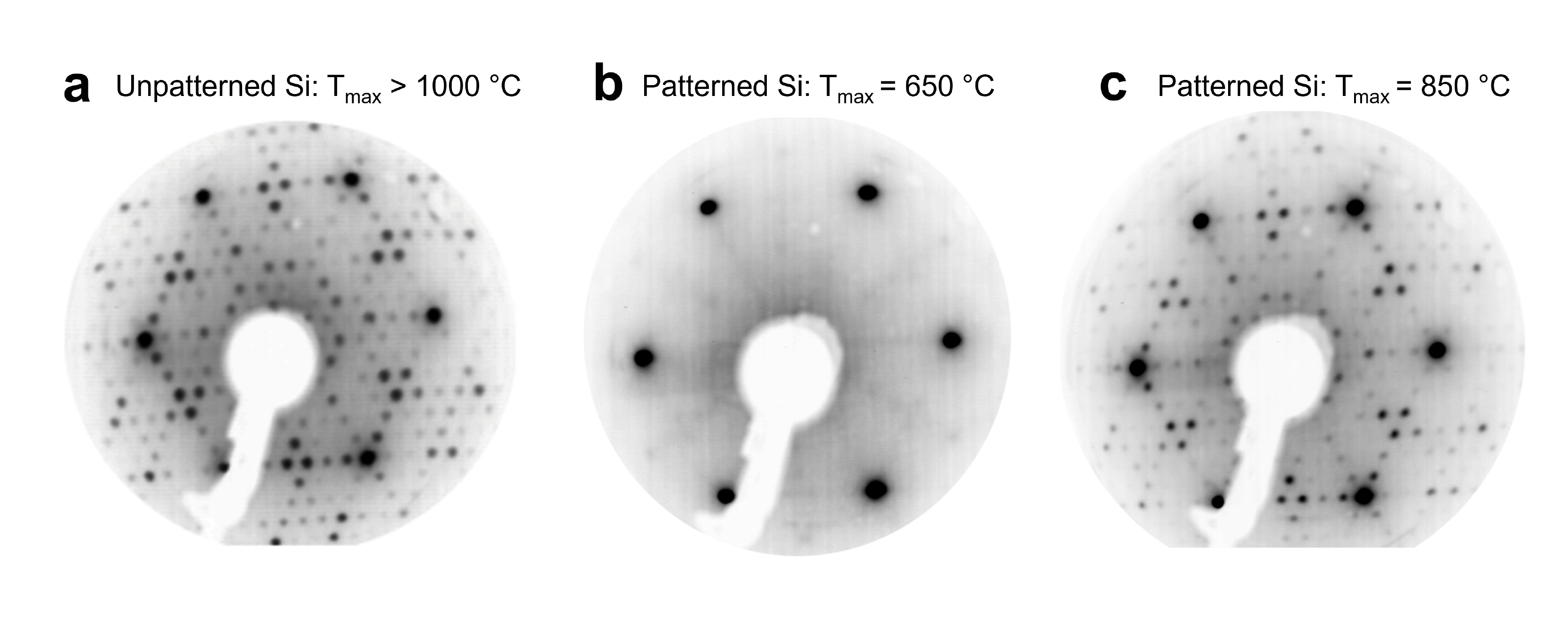} \par\end{centering}
\caption{LEED characterization of Si(111)$-(7\times7)$ surfaces. (a)~Diffraction from a fully $7\times7$ ordered surface obtained by successive heating cycles with a high maximum temperature of $>1000$~$^{\circ}$C, as used previously on unpatterned Si substrates \cite{paper335}. (b)~Incomplete $7\times7$ ordering (exhibiting only faint reconstruction spots) obtained by annealing a patterned H/Si(111) substrate at 650~$^{\circ}$C. Degreasing and HF etching are performed in a standard chemistry lab fume hood for this sample. The $7\times7$ pattern does not form as readily as in the previous samples. (c)~Improved $7\times7$ order obtained after successive heating cycles limited to a maximum temperature of $850$~$^{\circ}$C. The electron beam energy is 51~eV in all cases.}
\label{FigS2}
\end{figure*}

\begin{figure*}
\begin{centering} \includegraphics[width=0.9\textwidth]{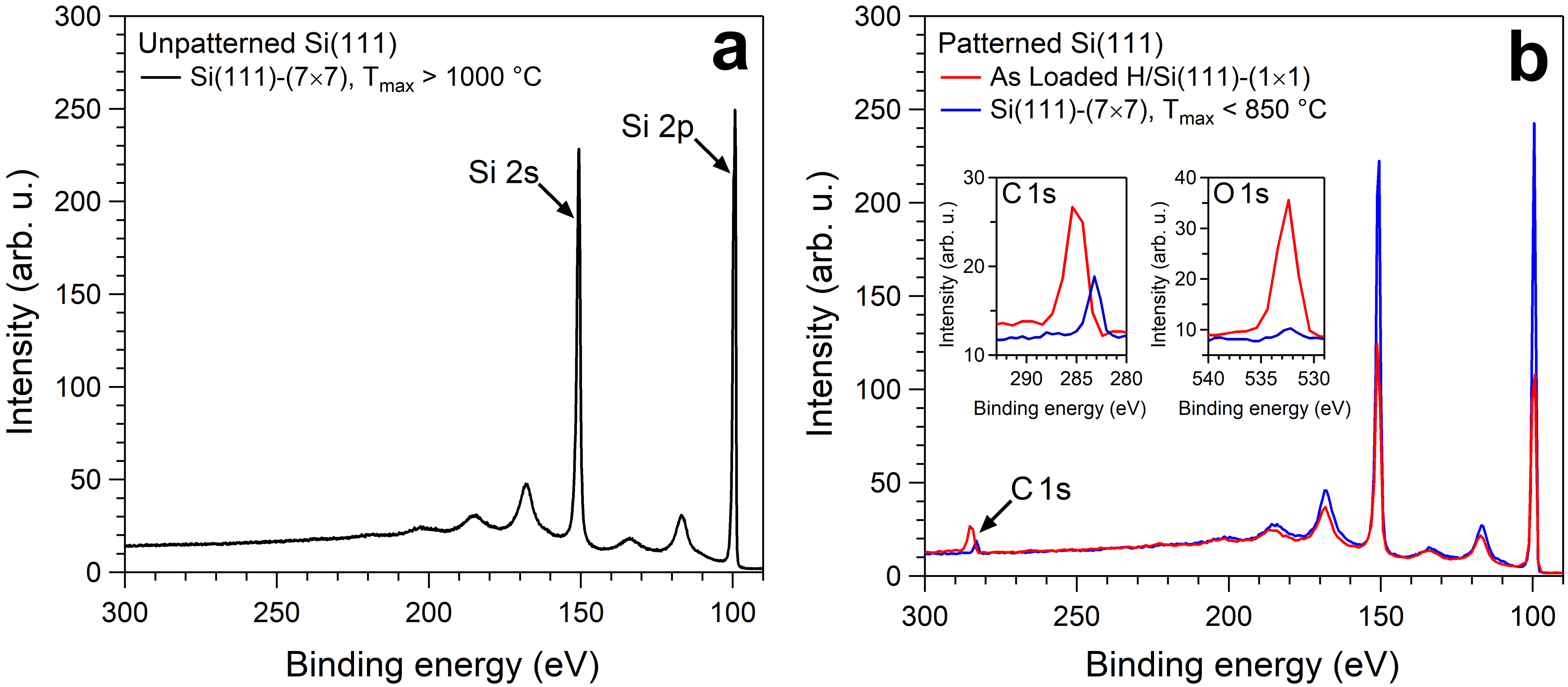} \par\end{centering}
\caption{X-ray photoemission spectroscopy (XPS) of (a) a clean Si(111)$-(7\times7)$ surface prepared on an unpatterned substrate following the method in Ref.~\cite{paper335} and (b) a patterned H/Si(111) substrate prepared as the described in Fig.~\ref{FigS2}. In (b), the spectra collected immediately after loading into an ultrahigh vacuum (UHV) chamber and after heating are plotted in red and blue, respectively. In all cases, the Si 2p and 2s core levels appear near 100 and 150~eV binding energy, respectively, each of them accompanied by a set of bulk-plasmon loss peaks, as well as a shoulder produced by the $\approx4$~eV chemical shift produced by a small remaining part of the native surface oxide \cite{H14_3}. The as-loaded H/Si(111) sample exhibits lower core-level peak intensities, a higher background, and some intensity appearing around 285~eV binding energy arising from C 1s, which is not detectable in (a). Labeled insets in (b) show the XPS signal in narrow ranges around the C 1s and O 1s core levels, which are clearly resolved. This is in contrast to Ref.~\cite{VES94}, where an extensively degreased, thick thermal oxide surface was used as a starting point for obtaining H/Si(111). Heating the sample in (b) nearly eliminates the oxygen contribution, although some carbon contamination remains with a chemical shift to $\approx283$~eV binding energy, consistent with SiC \cite{H14_3}. These remaining carbide impurities are likely the primary hindrance to $7\times7$ ordering, which limits the quality of the fabrication process compared with higher-temperature methods in which carbon is diffused into the bulk.}
\label{FigS3}
\end{figure*}

\begin{figure*}
\begin{centering} \includegraphics[width=0.9\textwidth]{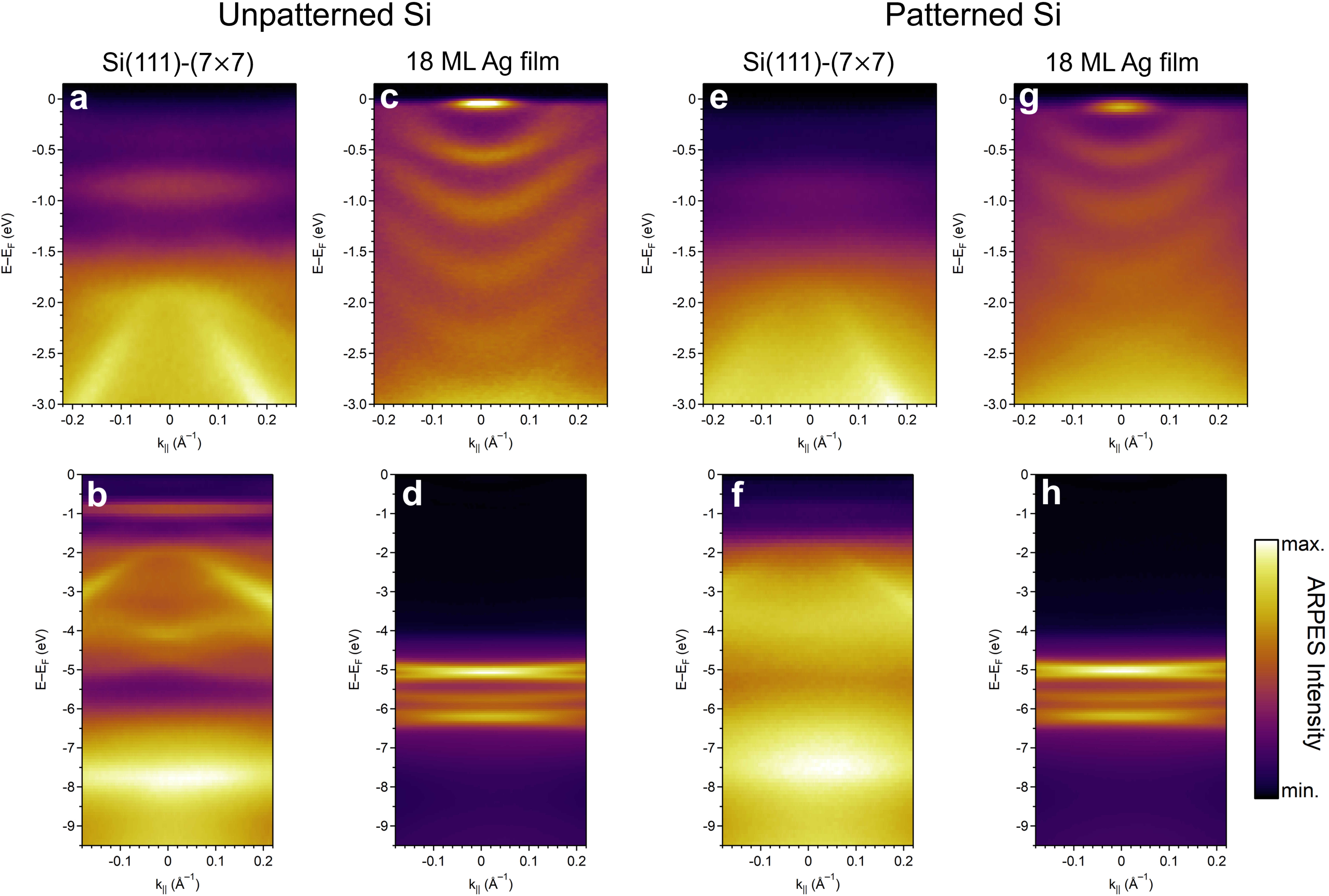} \par\end{centering}
\caption{Comparison of ARPES spectra collected from unpatterned (a-d) and patterned (e-h) substrates either before (a-b,e-f) or after (c-d,g-h) growth of 18~ML of Ag(111). We plot results for a narrow energy range within 3 eV of $\EF$ in panels (a,c,e,g) and over a wide energy sweep of the valence bands in panels (b,d,f,h). In general, we observe broader features increasing toward higher binding energies and a larger inelastic background in the patterned sample. The surface states of the $7\times7$ silicon surface around 0.25 and 0.8~eV below the Fermi level also show weaker intensity for the patterned sample. Both of these observations are consistent with the roughness imposed by etching most of the surface area of the substrate (away from the patterns), as well as a degradation off perfect long-range atomic ordering introduced by the presence of carbide, as discussed in Fig.~\ref{FigS3}.}
\label{FigS4}
\end{figure*}

\begin{figure*}
\begin{centering} \includegraphics[width=0.9\textwidth]{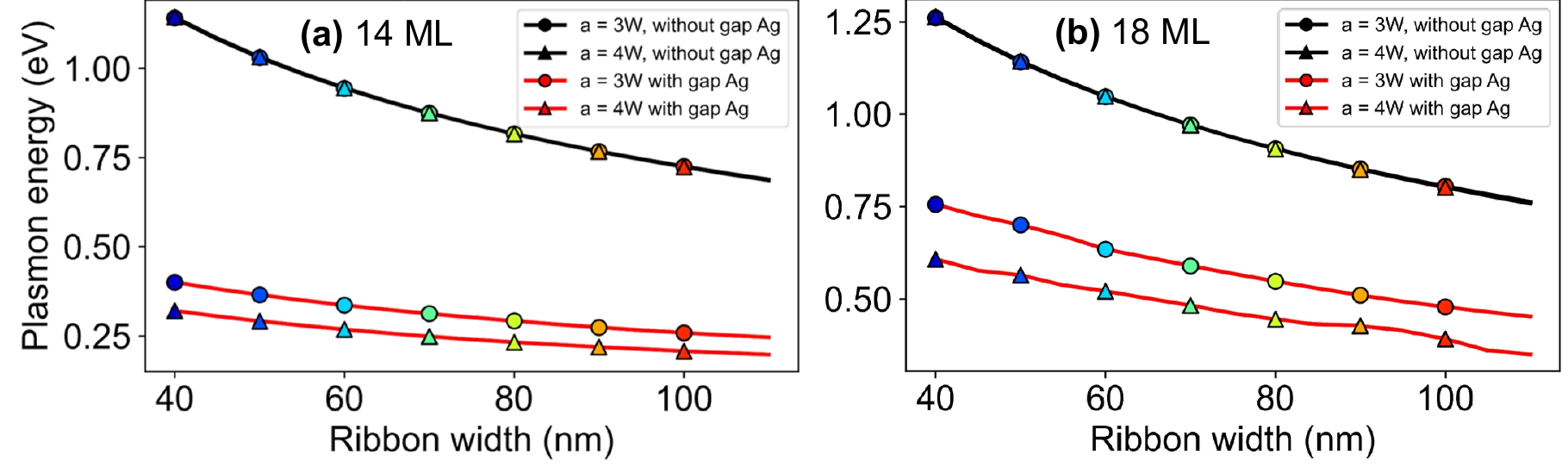} \par\end{centering}
\caption{Effect produced by silver deposition in the groove region on the ribbon plasmon energies. We show the plasmon energies calculated under the same conditions as in Fig.~1 in the main text for 14~ML (a) and 18~ML (b) silver films with a period-to-width ratio $a/W=3$ and 4 (see labels) and assuming the presence or absence of silver on the lower groove regions ({\it with} and {\it without gap Ag}, respectively). A substantial redshift is produced in the ribbon plasmon energies by interaction with the metal in the groove regions.}
\label{FigS5}
\end{figure*}


%

\end{document}